\documentclass{article}
\usepackage[T1]{fontenc}
\usepackage[utf8]{inputenc}
\usepackage[]{ismir} 
\usepackage{amsmath,cite,url}
\usepackage{amssymb}  
\usepackage{graphicx}
\usepackage{color}

\title{\vspace{-1mm}Music and Artificial Intelligence: artistic trends\vspace{-2mm}}

\oneauthor
  {{Jordi Pons, Zack Zukowski, Julian D. Parker, CJ Carr, Josiah Taylor, Zach Evans}\vspace{2mm}}
  {Stability AI\vspace{-5mm}}

\begin{document}

\maketitle

\begin{abstract}

\noindent We study how musicians use artificial intelligence (AI) across formats like singles, albums, performances, installations, voices, ballets, operas, or soundtracks. We collect 337 music artworks and categorize them based on AI usage: AI composition, co-composition, sound design, lyrics generation, and translation.
We find that AI is employed as a co-creative tool, as an artistic medium, and in live performances and installations. 
Innovative uses of AI include exploring uncanny aesthetics, multilingual and multigenre song releases, and new formats such as online installations.
This research provides a comprehensive overview of current AI music practices, offering insights into emerging artistic trends and the challenges faced by AI musicians.

\end{abstract}

\section{INTRODUCTION}

Recent progress in music AI has been driven by the success of text-to-music generators~\cite{agostinelli2023musiclm,evans2024long,evansfast}. Alongside raising ethical and copyright concerns~\cite{SUNO,UDIO,huang2023beyond,rezwana2023user,lee2022ethics},
these developments have also inspired the need for 
more controllable models~\cite{parker2024stemgen,nistal2024diff}, improved methodologies~\cite{batlletowards}, and open datasets~\cite{evans2024stable}.
Yet, one important question remains largely unaddressed: 
\textit{how are artists using artificial intelligence to make music?}
To answer this question, we collect a comprehensive set of AI music in section \ref{sec:method}, and analyze it quantitatively and qualitatively in sections \ref{sec:quantitative} and \ref{sec:qualitative}, respectively. In section~\ref{sec:discussion}, we also discuss and frame our findings within a broader historical and cultural context before concluding in section \ref{sec:conclusions}.

Previous research has already collected music to study AI musicians' practices. 
Huang et al.~\cite{huang2020ai} collected data from 13 entries participating in 2020's AI Song Contest and HAISP~\cite{Morris2024HAISP} collected data from 34 entries in 2024. 
The AI Song Contest explores AI’s potential for songwriting, since 2020.
Unlike previous research, we collect not only AI Song Contest entries, but also a broader range of AI music. As a result, we expand their scope by collecting 337 artworks that we release as a public dataset\footnote{https://github.com/jordipons/ai-music-artistic-trends}.
Literature reviews~\cite{civit2022systematic,nime2023_46}, surveys~\cite{fernandez2013ai,lopez2018algoritmic,herremans2017functional,papadopoulos1999ai}, and the AI Music Generation Challenge reports~\cite{sturm2022ai, sturm2020ai} 
are also closely related to our work, but their focus is on comparing AI music systems rather than studying musicians' practices.

\section{Scope and Methodology}
\label{sec:method}

We collect a comprehensive set of AI music works sourced from festivals, hackathons, artist websites, and social networks.
We also include all AI Song Contest entries.  We include artists who self-identify as such and engage in an artistic discourse. Works by casual creators (who use AI to make music for entertainment purposes) are excluded and discussed in section \ref{sec:casualcreator}. Our search is guided by the keywords: {artificial intelligence}, {deep learning}, and {music}. Only works with deep learning technologies are considered. Works not based on deep learning are excluded and discussed in section \ref{sec:algorithmicmusic}. This ensures the examination of contemporary AI music that aligns with state-of-the-art techniques. Hardware, plugins, and platforms are also excluded and discussed in section \ref{sec:musicAItools}. Technical demos are excluded and not discussed. We collect only works with evidence of AI use for music creation, excluding those that merely discuss AI or use it as a thematic element.
We aim to collect AI music published before July 31 (2025), and we define AI music as any musical artwork in which AI, particularly deep learning technologies, plays a role in supporting the music production process of an artist who self-identifies as such (not casual creators who use AI to make music for entertainment or experimentation purposes, without engaging in a broader artistic discourse).

For each AI music work, we collect the following data: project name, artist name, year of publication, release format, use of AI, and if it was part of the AI Song Contest.
During data collection, we identified various release formats: singles, albums, performances, (online) installations, AI voices, ballets, operas, and soundtracks. We also identified multiple uses of AI: AI-composition, co-composition, sound design, lyrics generation, and translation.
AI-composition identifies works with minimal artist intervention (\textit{e.g.}, text-to-audio or unconditional generation).
Co-composition works use AI for composing parts of the music (\textit{e.g.}, basslines, melodies, harmonies, or drum patterns).
Sound design is for works where AI is used to create sample packs, loops, synthetic vocals, or for audio-to-audio effects.
Lyrics generation is when AI is used to write song lyrics, while translation is when AI is used to translate lyrics into other languages and, optionally, to also sing them.
Data was collected and organized based on the categories above, followed by quantitative and qualitative analyses. 
Note that the above categories (\textit{i.e.}, the proposed ontology) are intended only to describe the collected works, as artistic practices may continue to evolve.
We publish the collected works as a public dataset$^1$. To further understand the collected data, we discuss an example: \textit{Relentless Doppelganger} by Dadabots (2019). This 24/7 livestream of AI-generated technical death metal qualifies as {AI-composition}, as is generated by an unconditional model, and is an {online installation} due to its livestream format. Finally, it was not part of the AI Song Contest.
Note the \textit{Relentless Doppelganger} example uses AI for AI-composition, but other examples may involve multiple AI uses (\textit{e.g.}, co-composition and sound design combined).

\section{QUANTITATIVE ANALYSIS}
\label{sec:quantitative}

\begin{figure}[t]
    \centering
    \vspace{-4mm}
    \includegraphics[width=\linewidth]{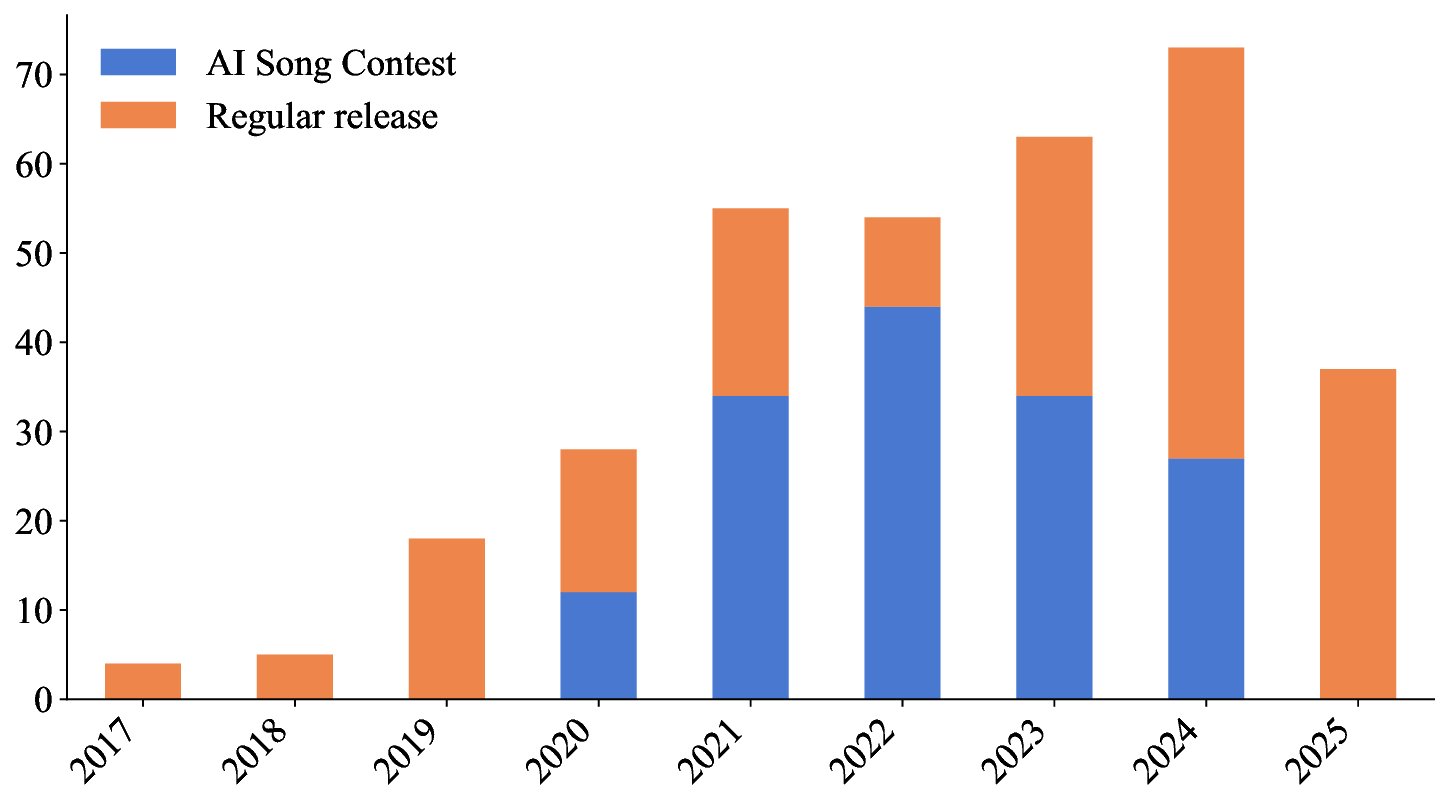}
    \vspace{-8mm} 
    \caption{\mbox{AI Song Contest vs Regular releases across years.}}
    \label{fig:one}
    \vspace{3mm} 
    \includegraphics[width=\linewidth]{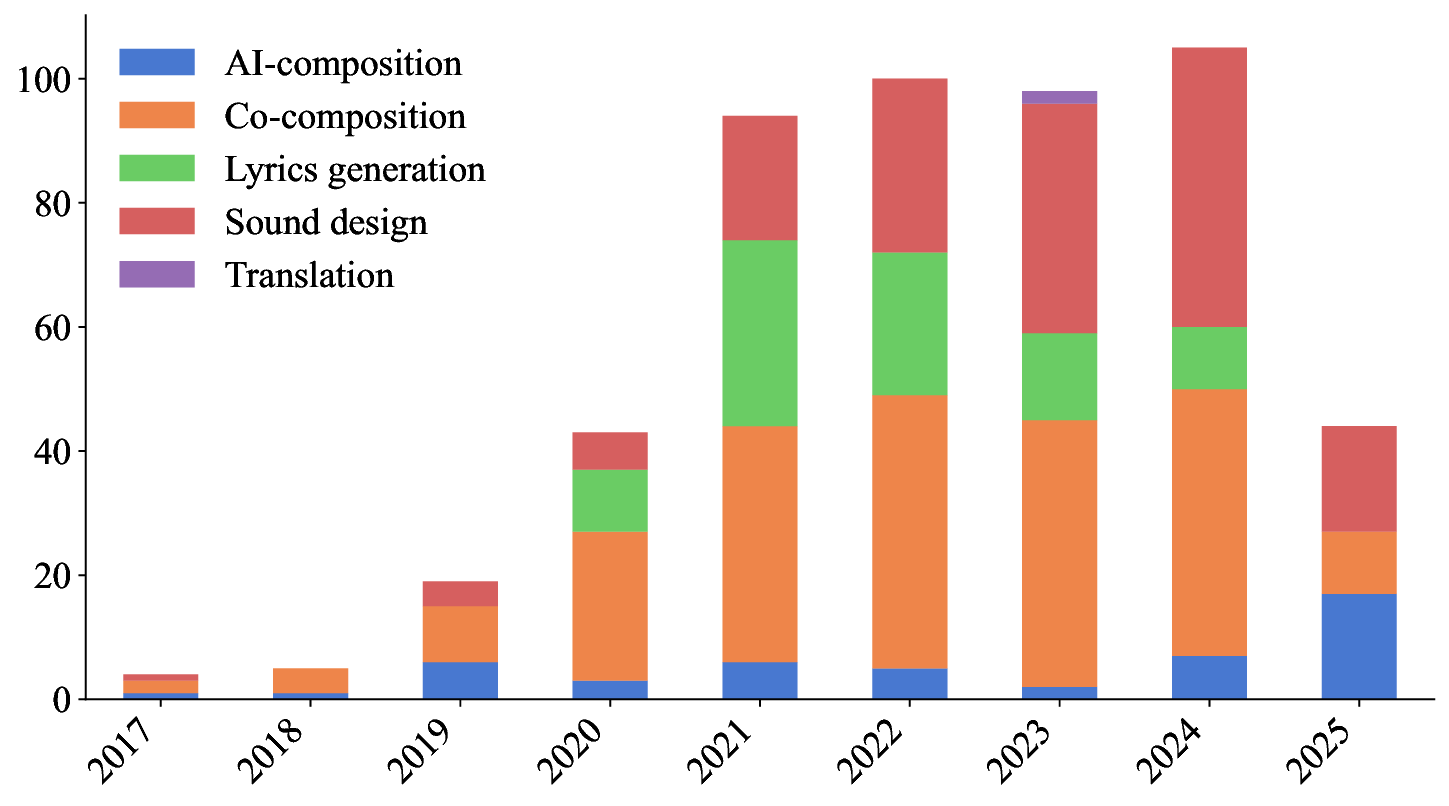}
    \vspace{-7.3mm} 
    \caption{Use of AI across years.}
    \label{fig:two}
        \vspace{-1mm}

\end{figure}

\begin{figure}[t]
    \centering
    \vspace{-3mm}
    \includegraphics[width=0.98\linewidth]{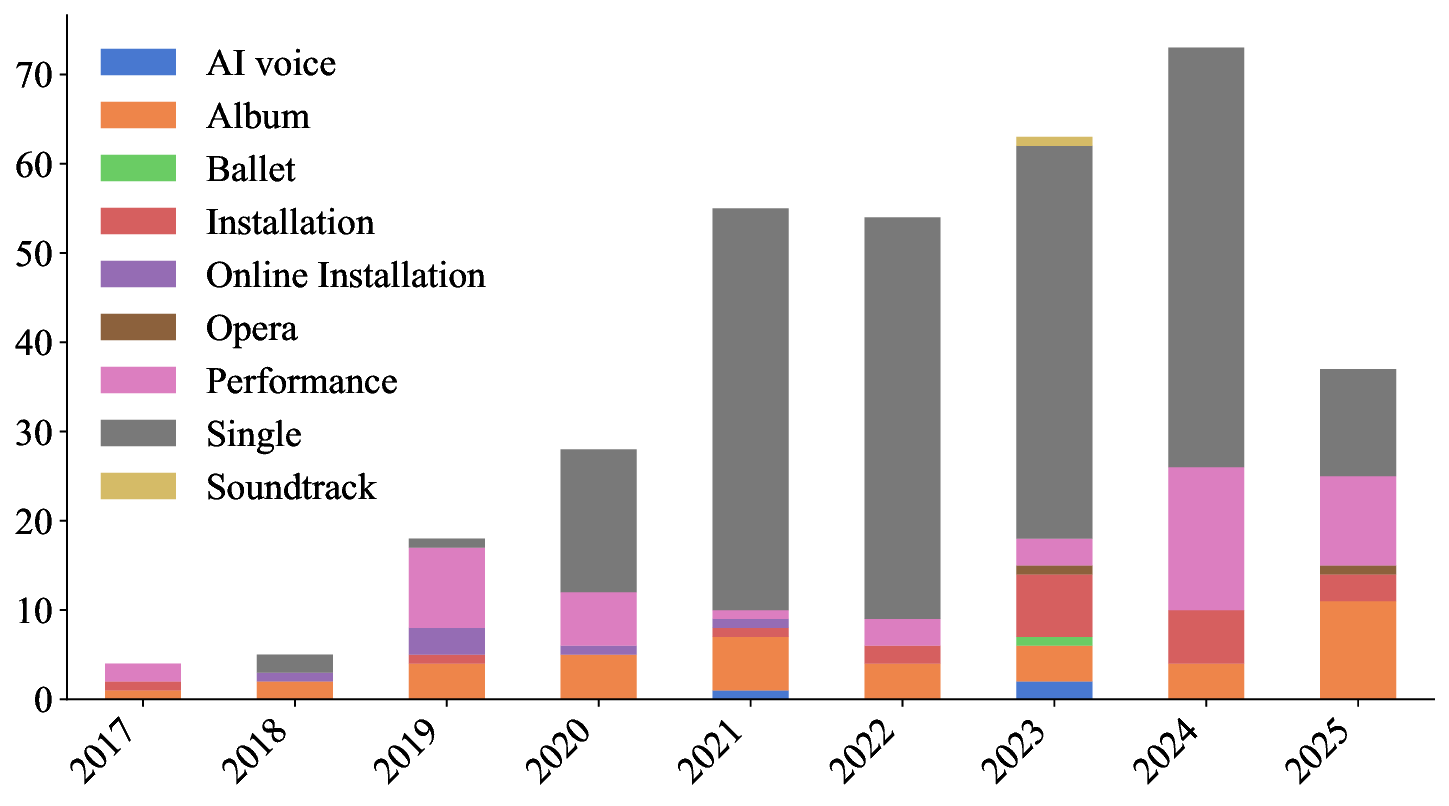}
    \vspace{-3mm}
    \caption{Format releases across years.}
    \label{fig:three}
    \vspace{5mm}
    \includegraphics[width=\linewidth]{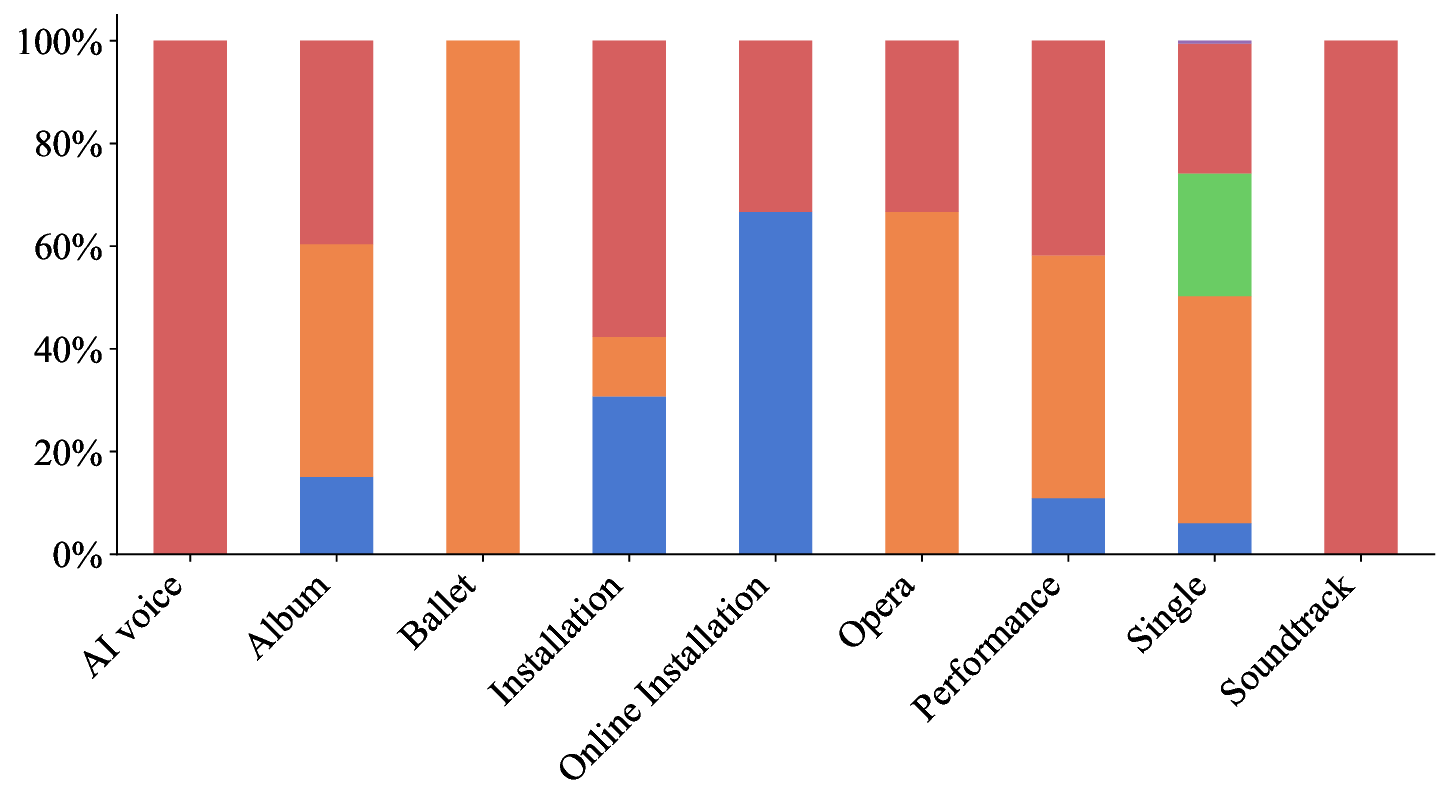}
    \vspace{-10mm}
    \caption{Use of AI across formats (legend in Figure 2).}
    \label{fig:four}
    \vspace{-1mm}
\end{figure}

We collected 337 music artworks and categorized them as described in section \ref{sec:method}. Figures 1 to 4 depict our findings. 
Figure \ref{fig:one} shows that the collected music extends well beyond previous research~\cite{huang2020ai,Morris2024HAISP}, which only studied AI Song Contest entries from 2020 and 2024. 
Figures~\ref{fig:two},~\ref{fig:three} and \ref{fig:four} show how AI has been used across years and formats. Most artworks are singles, performances, and albums. Further, AI is mostly used as a co-composition and sound design tool, but is also heavily used for lyrics generation.
Note, however, that the identified trends may be emphasized due to the absence of casual creators in our dataset (see section~\ref{sec:method}), who primarily engage in online text-to-music platforms for AI-composition \cite{UDIO,SUNO}.
Also, note that the \textit{y}-axis values in figure \ref{fig:two} differ from those in figures \ref{fig:one} and \ref{fig:three} as artworks can use AI in multiple ways, \textit{e.g.}, for lyrics generation and sound design. Also note that AI music production has been growing since 2017, though the volume of works has remained fairly steady over the past five years. Finally, the use of AI for lyrics generation is mostly for singles and has been decreasing since 2021.

We collected 212 singles (those do not include songs in albums). First, a large portion of singles were AI Song Contest entries. Second, only a few singles use AI for full-song generation (AI-composition, figure \ref{fig:four}). This might be caused because the artists (not casual creators) we study aim at preserving creative agency. Third, most singles used AI for sound design, lyrics generation, and co-composition (figure \ref{fig:four}). 
Finally, two singles were released multilingually with the help of AI (translation, figures \ref{fig:two} and \ref{fig:four}).

We collected 41 albums. Only a few were based on full-song generation (AI-composition, figure \ref{fig:four}). The rest used AI as a co-composition and sound design tool, which aligns with the trends observed in the collected singles.

A sizable portion of AI music works were performances and installations.
We collected 51 performances where AI models react live to human input for live co-composition, or as a live sound design tool that is often being performed as an instrument (see section \ref{sec:performances}). We also collected 27 installations, out of which 6 were online. 
Most installations use AI to generate music or soundscapes (AI-composition, figure \ref{fig:four}) or for interactive sound design (see section \ref{sec:installations}).

We collected 4 projects using AI as an artistic medium: 3 AI voices and 1 generative AI album (see section \ref{sec:medium}).

We also collected 2 operas (\textit{VALIS} and \textit{autoplay}), 1 soundtrack by Kensuke Ushio (\textit{Chainsaw Man}), and 1 ballet by Reeps One and Gadi Sassoon (\textit{Fusion}).

\section{QUALITATIVE ANALYSIS}
\label{sec:qualitative}

\subsection{Artistic agency}

Only few of the collected music works use AI-composition techniques. 
This is mainly because our dataset does not include casual creators and the studied musicians seek to retain artistic agency. 
By using AI as a co-composition or sound design tool, artists can maintain full creative control. This is in line with previous findings noting that AI musicians adopt `modular approaches' to `steer and assemble a conglomeration of models towards their creative goals'~\cite{huang2020ai}. For example, symbolic music models are common for generating ideas for melodies, harmonies, basslines, or drum patterns. Text-to-audio models are also used for curating samples and loops, or recent audio-to-audio capabilities are also employed for sound design. AI-assisted lyrics writing with chatbots is also becoming common among AI music songwriters. An early example is \textit{You Can't Take My Door} by Botnik Studios, but many AI Song Contest entries used AI for songwriting \cite{Morris2024HAISP}. Further, AI-assisted translation (and singing models) enable multilingual songs, expanding their reach internationally. Examples include MIDNATT’s \textit{Masquerade}, in English, Korean, Japanese, Chinese, Spanish, and Vietnamese; and Lauv’s \textit{Love U Like That}, both in English and Korean.

Yet, some artists intentionally explore the absence of artistic agency, critically and creatively embracing AI.
An early example is \textit{Relentless Doppelganger} by Dadabots~(2019), an infinite livestream of death metal generated from an unconditional model. A more recent example is \textit{The Laziness Complex} by Forest Little Music (2024), which the authors describe as `lazy artwork, lazy sound, lazy production, and a lazy, relaxing mood'. AI-generated music provides artists with a means to engage with and critically examine AI, allowing them to challenge prevailing narratives and contribute to their reinterpretation.

\subsection{How does AI music sound?}

AI music sound can be nearly indistinguishable from non-AI music, but it can also embrace the flaws and aesthetics of AI.
On one hand, works such as those by SKAGGE, YACHT or Taryn Southern sound nearly indistinguishable from non-AI music, requiring listeners to know in advance whether AI was used for this production or not. One advantage of this approach is that it uses familiar musical elements, making it accessible to wider audiences.
On the other hand, AI music can embrace AI's flaws and aesthetics, making it challenging for broader audiences.
This is exemplified by the early works of Damian Dziwis, Dropcontrol, or Hiroshi Yamato (2017-2019), but this experimental trend continues from 2020 to 2025. 
GANTASMO and Caramel Freeman cultivated the uncanny \cite{Tokui2024} as a way to strike a balance between embracing the aesthetics of AI and using familiar music elements. The uncanny describes things that are familiar yet strange or unsettling.
To that end, they use synthetic voices and AI artifacts to create a noticeably uncanny, AI-driven sound, with rhythms and song structures that are familiar to mainstream audiences.
Holly Herndon’s album \textit{PROTO} also explored the uncanny. Further, many past artists explored the uncanny, including surrealists, symbolists like Redon, and romantic painters like Goya. 
Current AI music often evokes an uncanny feeling due to its imperfections and, like past artists, AI musicians can explore the uncanny to reflect modern struggles.

Dadabots also explored multi-genre music \cite{dadabotsmultigenre}. As AI can generate music in multiple genres, this can also be a compelling {new} AI sound.
Before, musicians had to dedicate years to learn a specific genre. Now, AI can blend and produce music in multiple (and potentially new) genres.

The evolving legal landscape \cite{SUNO,UDIO} around training datasets has driven artists to become more deliberate in sourcing their own training data. As such, artists are curating datasets for custom training to craft their unique sound, much like sound designers program synthesizers. Examples include Rob Clouth’s \textit{Cheatboxer Jams}, allowing intricate AI sound design with his organic voice control, or Reeps One’s \textit{Second Self}, a generative beatboxing model trained on his performances. Emptyset (\textit{Blossoms}) used a collection of their existing material as well as 10 hours of improvised recordings using wood, metal and drum skins.

\subsection{Generative AI as an artistic medium}
\label{sec:medium}

One significant trend is the novel use of AI as an artistic medium for releasing AI voices and generative albums.

Artists like Holly Herndon, Grimes, and Sevdaliza have released AI models of their voices, allowing others to create music using their vocal likenesses. This development introduces a new degree of interaction between artists and their audiences, enabling collaborations that transcend traditional creative boundaries. In all cases, revenue-sharing models have been introduced, requiring around a 50\% split on royalties for tracks using their AI voices. Beyond artistic expression, these developments raise novel questions about digital identity and authorship. Holly Herndon proposed a governance model where token holders vote on how her AI voice is used, distributing ownership and decision-making among a decentralized community. 

Expanding beyond voice models, SENAIDA released \textit{Kunst Kaputt}, a collaborative generative AI album where listeners can use AI to remix her tracks and release them, with profits shared equally between remix creators and SENAIDA.
Critiquing streaming platforms as 
`broken', her initiative challenges current distribution and business models by prioritizing collaboration with fans. She encourages artists to rethink how emerging technologies shape their creative expression and economic sustainability.
Note the similarity between \textit{Kunst Kaputt}'s AI album and Eno’s generative music album (section \ref{sec:generativemusic}). This generative perspective, whether AI-driven or not, turns music into a dynamic/ephemeral medium that evolves with each listen.

\subsection{Impact in popular culture}
\label{sec:popular}

AI is also increasingly influencing mainstream music. One example is \textit{Heart on My Sleeve} by Ghostwriter, featuring AI vocals of Drake and The Weeknd, that raised concerns related to the reproduction of those artists' voices. 
Following this controversy, Drake’s diss track targeting Kendrick Lamar, titled \textit{Taylor Made Freestyle}, was removed from streaming platforms due to the use of AI vocals that mimic Tupac and Snoop Dogg. 
Music producer Metro Boomin released another diss track, \textit{BBL Drizzy}, based on an AI-generated song created with Udio by King Willonius. The track was used as the starting point for a diss track contest targeting Drake. 
Drake responded by rapping over \textit{BBL Drizzy} weeks later in \textit{U My Everything}, with Sexyy Red. King Willonius (not Metro Boomin) is listed among the songwriters for \textit{U My Everything}. 
These examples reflect AI's growing influence on hip hop and popular culture.

AI played a crucial role in the Grammy-recognized production of \textit{Now and Then} by The Beatles, where AI techniques were used to restore John Lennon's voice from an old demo. 
Similarly, Jane `Nightbirde' Marczewski’s family released \textit{Still Got Dreams} three years after her death. 
Another example is PAMP!'s tribute \textit{Florespiña} to the Galician singer Ana Kiro (1942–2010). PAMP! also has the support of the family.
Finally, AI-generated vocals allowed country singer Randy Travis, who lost his voice after a stroke, to release his first new song in over a decade.

Finally, AI-generated music is attracting substantial audiences online. On Spotify, the pop-rock project \textit{The Velvet Sundown}, an `artistic provocation designed to challenge\,(...)\,authorship, identity, and the future of music itself in the age of AI', has obtained millions of streams. On YouTube, \textit{Brainrot Italianini} has gained millions of views with its absurd, meme-driven style, exemplifying AI music that embraces absurdism and parody. On Suno, artists like \textit{Feline Music} (brainrot), \textit{Moisty} (Japanese lo-fi), and \textit{ImOliver} (indie-pop) have also reached millions of plays.

\subsection{Performances}
\label{sec:performances}

Many AI-based music works are designed to be performed live, \textit{e.g.}, Nina Masuelli performed \textit{The Jar}, an interactive AI instrument, as part of Tod Machover’s opera \textit{VALIS}.

Further, a common structural feature in AI-based performances is the use of the question–answer format. This is exemplified in \textit{Algo-Rhythm} by Keisuke Nohara and Ryosuke Nakajima, which features a live drumming exchange between a drummer and an AI. A similar interaction is seen in \textit{Piano + AI} by Marco Mezquida, where the pianist engages in a call-and-response dynamic with the AI in a jazz improvisation context. In \textit{Second Self}, beatboxer Reeps One engages in a live musical dialogue with an AI, effectively staging a human–machine beatbox battle.

Artists are exploring novel computer-based interfaces. Manaswi Mishra’s \textit{Algorave} or performances with Holly+ employ voice-based controls. Nicola Privato’s \textit{Mouja} combines magnets and AI in real-time. The Flaming Lips’ \textit{Fruit Genie} integrates \textit{Piano Genie}~\cite{donahue2019piano} with a fruit-based physical interface. AI models trained on Israel Galván’s flamenco footwork (\textit{sapateado}) performed alongside him, with rhythms sonified by custom robots. \textit{Beatbots} by Pu et al. \cite{pu2025beatbots} uses multiple robots to create complex and unconventional music beyond human capabilities.

Another area of exploration is the intersection of AI and acoustic music. Early examples (in 2019) include AIVA's AI compositions performed by an orchestra; or \textit{Alter} by Robert Laidlow and PRiSM, which use AI as a collaborative and interactive tool performed by musicians on stage. Dziwis et al.~\cite{dziwis2014conductor} developed a gesture-controlled system that combines piano with AI music and real-time visuals.

AI DJ sets are also popular. Early examples include Nao Takui’s \textit{AI DJ} (2017) and Peter Kirn’s techno performance at the GAMMA Festival (2019). Recent examples include Dadabots' \textit{Prompt Jockeys}, and Google's \textit{MusicFX DJ}.

\subsection{Installations}
\label{sec:installations}

AI music is also exhibited as an installation, what allows exploring AI’s potential for real-time music generation and interactive sound design.
Artists are leveraging AI’s generative capabilities to create installations that react dynamically to audience presence. 
In \textit{TECHNE: The Vivid Unknown}, John Fitzgerald integrates generative visuals and music that respond to visitor movement. Similarly, \textit{Transformirror} by Daito Manabe and Kyle McDonald uses AI to both transform visitors’ images and produce background music. Or \textit{Junkyard RAVE} by Manaswi Mishra is an interactive sculpture that explores the latent space in real-time. Further, AI-generated soundscapes play a central role in Yaboi Hanoi’s \textit{This is not Bird Song} and Refik Anadol’s \textit{Living Architecture: Gehry}. Hence, some installations include not only AI-generated music but also AI-generated soundscapes and sound effects. Finally, Holly Herndon and Mat Dryhurst’s \textit{The Call} installation explores the role of training data: “if all media is training data, including art, let’s turn the production of training data into art instead”.

A notable part (6 of 27) of the installations were online. Examples include \textit{Relentless Doppelganger} by Dadabots, a continuous livestream of AI-generated death metal, or interactive music experiments such as \textit{The Incredible Musical Spinners from Latent Space} by Tero Perviainen or Google’s \textit{AI Bach Doodle}, which enable users to generate short AI-driven compositions in various styles~\cite{huang2019bach}. Other examples include \textit{The Machine Folk Session}, a website with folk music generated by (or co-created with) AI, and interactive instruments like Qosmo’s \textit{Neural Beatbox}. Due to AI’s digital nature, there exists a trend toward online interfaces that enable user interaction. This online trend also reflects the constraints imposed by the COVID-19 pandemic, which coincided with this study’s timeframe.

\subsection{Year-by-year analysis}

This section chronologically highlights a few works that, based on our analysis, stand out for cultural impact. A full chronology is available online through our dataset release$^1$.

\textbf{\textit{2017}} --- Nao Tokui used deep learning for \textit{AI DJ}, and Damian Dziwis and Dadabots used AI to make experimental music in symbolic and audio domains, respectively.

\textbf{\textit{2018}} --- HANZ and Taryn Southern produced music from AI-generated symbolic music, while Botnik Studios used AI-generated lyrics to that end. 
But AI music was also experimental, as in \textit{GIVE WAY} by dropcontrol and Hiroshi Yamato. 
Tero Parviainen built an online installation.

 \textbf{\textit{2019}} --- AI and acoustic music was combined. Those key projects were presented: \textit{Google's Bach Doodle}, \textit{Relentless Doppelganger} (Dadabots), \textit{Machine Folk} (Bob Sturm), \textit{Second Self} (Reeps One), \textit{PROTO} (Holly Herndon), \textit{Chain Tripping} (YACHT), and \textit{Blossoms} (Emptyset). 

 \textbf{\textit{2020}} --- The AI Song Contest kicks off this year. Experimental music was produced, like \textit{Transfiguración} (Hexorcismos). The single \textit{Sinatra Sings Britney Spears} (Dadabots) was released.
 Two big productions in Russia (Artypical) and in Japan (Israel Galván, YCAM, Nao Tokui).

 \textbf{\textit{2021}} --- \textit{Holly+} and \textit{UNICA} (DeLaurentis) were released. Never Before Heard Sounds presented the \textit{Features Music} single and the installation \textit{GAN.STYLE}. 

 \textbf{\textit{2022}} --- Multicultural AI music was introduced by PAMP! and Yaboi Hanoi. \textit{WIRE} (PortraitXO) and \textit{I Stand in the Library} (Ed Newton-Rex) were also presented.

 \textbf{\textit{2023}} --- The ballet \textit{Fusion} (Reeps One, Gadi Sassoon), \textit{Chainsaw Man}'s soundtrack (Kensuke Ushio), and the opera \textit{VALIS} (Tod Machover) were released, as well as two AI voices: \textit{Dahlia} and \textit{GrimesAI}.
Midnatt and Lauv released 2 multilingual songs. \textit{Models} (Lee Gamble), \textit{Now and then} (Beatles), and \textit{Heart on My Sleeve} (Ghostwriter w/ AI Drake and The Weeknd voices) were also released.

 \textbf{\textit{2024}} --- Drake's controversy continued: \textit{Taylor Made Freestyle} (Drake) and \textit{BBL Drizzy} (King Willonius, Metro Boomin). \textit{Prompt Jockeys} \& \textit{MusicFX DJ} were presented.

 \textbf{\textit{2025}} \textit{(until July)} --- SENAIDA released \textit{Kunst Kaputt}, exploring the potential of generative AI albums.
AI-generated music (\textit{e.g.}, \textit{The Velvet Sundown} or \textit{Brainrot Italianini}) starts to attract large audiences on online platforms.

\vspace{-1mm}

\section{Discussion}
\label{sec:discussion}

We further discuss our findings and contextualize them within a broader historical and cultural framework. After all, what is art without its historical and cultural context? In Table 1 we summarize the main examples we discuss.

\subsection{Artistic impact of emerging technologies}
\label{sec:artimpact}
Technological advancements have played an important role in shaping artistic expression across various forms. For example, the camera obscura influenced the development of realism in painting as it allowed tracing projected scenes onto canvas, better capturing perspective and realism. 
Also, the daguerreotype transformed portraiture. The daguerrotype was the first publicly available photographic process, and it challenged portraiture by making it cheaper, faster, and more precise than hiring a painter. 
As industrialization and mass production grew, so did its impact on the art world. The work of artists like Duchamp and Warhol exemplifies how industrial processes and mass production influenced modern art movements.
Similarly, in the realm of music, the introduction of the iron frame in pianos enabled composers like Beethoven to push the boundaries of musical expression~\cite{rehfeldt2021beethoven}. It allowed increasing string tension, expanding the potential of the piano both in terms of sound and dynamics~\cite{rehfeldt2021beethoven}.
Further, the introduction of amplifiers facilitated the rise of genres like rock, blues, and jazz. Amplifiers also led to the emergence of solo artists and singer-songwriters who could now play in large venues. Additionally, the use of tape-based samplers enabled musicians such as The Beatles and David Bowie to explore new sounds. Tape-based samplers, like the chamberlin or the mellotron, are keyboard instruments with tape loops that play recorded sounds.
Also, the development of analog synthesizers further expanded sonic possibilities, influencing genres ranging from electronic music to progressive rock.
Finally, drum machines, modern samplers, and the auto-tune were key technological innovations that influenced the development of new music genres. For example, the TR-909 drum machine influenced electronic dance music genres such as techno, house and acid house. Or the Akai MPC sampler had a major influence on the development of electronic music and hip hop.

As AI continues to develop, its influence on music is becoming noticeable (section \ref{sec:popular}) but the full extent of its impact on popular music and culture remains uncertain.

\begin{table}[]
\centering
\begin{tabular}{l|ccccc}
                   & \rotatebox{90}{Artistic impact} &  \rotatebox{90}{Cultural pushback} & \rotatebox{90}{Commodification of art} \\ \hline
Camera obscura \hspace{1.7mm} \textit{\small(photography)}    & \checkmark &  \checkmark   &                     \\
Piano's iron frame & \checkmark &                &                      \\
Daguerrotype \hspace{4.7mm} \textit{\small(photography)}     & \checkmark & \checkmark &   \checkmark   \\
Radio &            & \checkmark & \checkmark      \\
Amplifier & \checkmark  &  &       \\
Sampler & \checkmark & \checkmark      &    \\
Analog synthesizer& \checkmark & \checkmark          &     \\
Drum machine     & \checkmark & \checkmark       &      \\
Generative music     & \checkmark &        &     \\
DAW (digital audio workstation)               &  &                  &  \checkmark \\
Auto-tune          & \checkmark & \checkmark       &      \checkmark        \\ 
Artificial Intelligence          & ? & \checkmark     &  \checkmark     \\ 
\end{tabular}
\caption{Examples discussed in section \ref{sec:discussion}. The ? denotes unknown, as the artistic impact of AI on music history remains unclear at this early stage. In chronological order.}
\label{tab:thematic}
\vspace{-4mm}
\end{table}

\subsection{Cultural pushback}
\label{sec:pushback}
Throughout history, the artistic merit of certain works has been widely debated.
Vergnaud~\cite{Vergnaud1828}, an art critic, complained about the extensive use of camera obscura in paintings at the 1827 Salon exhibition in Paris.
Or Baudelaire, a strong critic of early photography, questioned the artistic legitimacy of the daguerreotype photographs~\cite{arago1839rapport,Baudelaire1859}. Interestingly, he himself was later immortalized through daguerreotype portraits, illustrating how new media can transition from controversy to cultural acceptance.
Duchamp’s \textit{Fountain} also clashed with prevailing artistic norms, facing significant resistance \cite{sloan1917}. Likewise, Warhol’s \textit{Campbell’s Soup Cans} were initially met with skepticism \cite{contemporaryartissue_2024}.

Music has witnessed similar controversies. In the early days of the radio, live performances were broadcasted. Recorded music was only later introduced, facing resistance from various stakeholders~\cite{landesberg2006rise}: broadcasters feared that recorded music would drive listeners away~\cite{rasmussen2008lonely}; the recording industry saw free broadcasts as an infringement on its rights~\cite{rasmussen2008lonely}; musicians' unions organized to resist the playing of records over the air~\cite{rasmussen2008lonely}; or networks labeled recordings as `canned' music and often prohibited broadcasting discs~\cite{rasmussen2008lonely}.
Notably, 1940s court cases established that prohibitions against recorded music had no legal basis~\cite{rasmussen2008lonely}. 
Further, the use of samples was once denounced as artistic `theft' rather than genuine creativity~\cite{evans2010sampling,Runtagh2016,olabode2023sampling}. 
Likewise, auto-tune users faced criticism for altering the natural quality of the voice, often viewed as compromising vocal authenticity. 
The initial reception of analog synthesizers was also mixed. 
Queen even stated in their album liner notes that they did not use synthesizers \cite{griffiths2023queen} to highlight the sounds created by Brian May with his guitar.
These cases underscore the evolving nature of artistic evaluation, where initial skepticism can transform into historical validation and cultural significance.

As new technologies gain the general public's attention, polarized opinions tend to arise. Eco~\cite{eco1964apocalittici} studied the relationship between mass culture and intellectual critique, distinguishing opposing trends: the \textit{apocalyptic} and the \textit{integrated}. Interestingly, the \textit{apocalyptic} narratives around the daguerreotype, the radio, and Duchamp's work closely resemble those around AI today~\cite{arago1839rapport,Baudelaire1859, claudio, rasmussen2008lonely,sloan1917}. 
Finally, the challenges associated with emerging technologies offer artists an opportunity to engage with and critically explore those. Artists can use AI creatively as a new form of expression, just as the auto-tune or sampling, but can also adopt AI as a new artistic medium, just as the daguerreotype was. In doing so, artists not only can challenge dominant narratives but also can help reshape them.

The daguerreotype disrupted portrait painters~\cite{arago1839rapport,Baudelaire1859};
live musicians playing on air were displaced as radio turned to recorded music\cite{rasmussen2008lonely}; synthesizers, samplers and drum machines reduced the demand for session musicians~\cite{jones2020sweet,Davies2020, RollingStone1969AFM}; and digital audio workstations (DAWs) reduced reliance on hardware engineers as software became available.
It remains unclear how AI will affect musicians in the long term, as skills and trends continue to develop.

To close, the evolving public perception of AI tools can also influence artistic output. For example, the AI Song Contest 2024 organizers recommended `against the use of AI tools Suno and Udio due to pending litigation'~\cite{SUNO, UDIO}. Hence, AI music artworks are the result of both technological progress and society’s response to these new tools.

\subsection{Casual creators and the commodification of art}
\label{sec:casualcreator}

Music generation tools~\cite{agostinelli2023musiclm,evans2024long,evansfast} today make song composition effortless, requiring no formal training or technical expertise. This shift is redifining musicianship, giving rise to casual music creators and the commodification of music. 
Casual creators use AI music tools for entertainment or experimentation purposes, without engaging in a broader artistic discourse. 
Further, Udio reported generating millions of tracks per week \cite{UDIO} and, based on Deezer's data, around 10\% of daily uploads to streaming services are AI-generated tracks (AI-composed, see section \ref{sec:method})~\cite{bernet2025deezer}. 
As AI continues to commodify music creation, concerns arise: is the sheer volume of AI-generated content overshadowing artists? is this content driving musical innovation or is it ephemeral entertainment? 
That said, studying casual creators' work is beyond our scope (see sections \ref{sec:method} and \ref{sec:limitations}).

Duchamp’s \textit{Fountain} also challenged established art, 
encountering opposition from the Society of Independent Artists: `the so-called {readymade} philosophy (...) exploits actual painters and sculptors, using their exhibition space without justification or craftsmanship' or it `incentivizes the mass replacement of human artistic labor with commercially available bathroom fixtures'~\cite{sloan1917}.
Just as Duchamp challenged established norms, certain AI music artists and platforms are challenging today’s music industry norms (see sections \ref{sec:medium}, \ref{sec:popular} or \cite{SUNO,UDIO}). 
Furthermore, Warhol embraced mass production. Using mechanical techniques like silkscreen printing he challenged the distinction between art and consumer products, effectively commodifying art.
Similarly, current AI music tools~\cite{parker2024stemgen,nistal2024diff,agostinelli2023musiclm,evans2024long,evansfast} allow anyone creating music with ease. 
This illustrates a shift where creativity can separate from its material realisation.
Caramiaux et al.~\cite{caramiaux2025generative} argue that this separation could be problematic as it withdraws the skills traditionally needed in creative processes. Yet, it can be an opportunity to build a nuanced artistic discourse, much like Duchamp and Warhol did.
The commodification of art can also be viewed through the lens of \textit{high} and \textit{low} art~\cite{fisher2013high}.
Bourdieu~\cite{pierre1979distinction} argues that social class influences aesthetic preferences, and `cultural capital' determines whether art is considered \textit{high} or \textit{low}.
Greenberg~\cite{greenberg1939avantgarde} argued that kitsch (\textit{low}) art is mass-produced and a simulacrum of avant-garde (\textit{high}) art that resists consumerism.
Adorno~\cite{adorno1944culture} also criticized popular music for its standardization and commercialism, calling for a shift towards avant-garde music.
Benjamin~\cite{benjamin1936work} notes that mass reproduction devalues the uniqueness of \textit{high} art, offering a chance to challenge existing power structures.
Levine~\cite{levine1990highbrow} examined how cultural hierarchies develop and found that forms of popular culture can achieve elite status. This illustrates the fluidity between what is considered highbrow and lowbrow art, using his terminology~\cite{levine1990highbrow}.

\subsection{Algorithmic music}
\label{sec:algorithmicmusic}

Algorithmic music refers to compositions created using rules or procedures. 
It has a rich history dating back to the Middle Ages, beginning with compositional techniques such as the isorhythm~\cite{bent2001isorhythm}. This technique pairs rhythmic and melodic patterns, that are often of unequal lengths, producing evolving textures as they cycle.
In the 17th century, Kepler’s \textit{Harmonices Mundi} linked planetary motion to musical intervals, and Kircher’s \textit{Arca Musarithmica} introduced a combinatorial device to generate music algorithmically.
During the 18th century, composers played dice games to generate music, using dice rolls to determine the sequence of prewritten musical bars~\cite{nierhaus2009algorithmic}.
The 20th century introduced serialism, a compositional method that began primarily with Schoenberg's 12 tone technique that arranges 12 pitches into a specific sequence that composers follow (in order) before any pitch can be repeated. This method ensures that no single pitch dominates, avoiding tonal hierarchies. Stockhausen and Boulez expanded the idea into total serialism, applying serialization not only to pitch but also to rhythm, dynamics, and articulation.

Computer music was also introduced during the 20th century.
One of the first examples of computer-composed music was the \textit{Illiac Suite} (1957) by Hiller and Isaacson. It has four movements, each exploring different ideas around melody, four-voice harmony, rhythm/dynamics, and probabilistic models.
Spiegel also pioneered computer music by creating \textit{Music Mouse} (1986), a computer-based musical instrument that allows creating expressive music without needing traditional skills in music theory or notation, by providing intuitive control over harmony and melody.
Spiegel composed several works using \textit{Music Mouse}, like \textit{Three Sonic Spaces} (1989). Early AI music efforts also date back to the 1980s, with David Cope’s \textit{Experiments in Musical Intelligence} generating music in the style of composers like Bartók or Chopin~\cite{cope1991computers}.
During the 21st century, quantum computer music has emerged as an area of exploration, involving the use of quantum computing principles to generate music~\cite{miranda2022quantum}.
At the same time, novel methods have drawn increased attention to AI-driven algorithmic music composition~\cite{cope1991computers,briot2020deep}. A notable example is \textit{Hello World} (2019) by Benoît Carré and François Pachet, who focused on generating pop music with AI~\cite{briot2020deep}.

Algorithmic music techniques include grammars, symbolic methods, knowledge-based systems, Markov chains, artificial neural networks, evolutionary algorithms, cellular automata, and hybrid systems \cite{fernandez2013ai, papadopoulos1999ai,maurer1999brief,briot2020deep,miranda2013readings,roads1996computer,schwanauer1993machine,todd1991music,nierhaus2009algorithmic}.
Further, such techniques can be stochastic (generative) or not \cite{maurer1999brief}.
In this section, we introduced algorithmic (and AI) music that does \textit{not} rely on deep learning, providing context for the music in sections \ref{sec:method}, \ref{sec:quantitative}, \ref{sec:qualitative}  ({relying} on deep learning). In section \ref{sec:generativemusic}, we discuss generative music (stochastic algorithmic music) that may or may not rely on deep learning.

\subsection{Generative music}
\label{sec:generativemusic}

Generative music, created through stochastic processes, existed long before generative AI. Often referred to as indeterminate, chance, stochastic, or aleatoric music, generative music embraces randomness either in composition or performance.
Early generative music can be traced to the 17th century with instruments like the aeolian harp, which creates evolving harmonic sounds through the unpredictable movement of wind~\cite{White_EerieInstruments}. In the 18th century, games like the \textit{Musikalisches Würfelspiel} (musical dice game) used dice rolls to sequence prewritten musical bars~\cite{nierhaus2009algorithmic}.
The 20th century brought significant innovation, as composers challenged traditional musical structures. Marcel Duchamp’s \textit{Erratum Musical} (1913) used shuffled slips of paper to determine the pitch sequence for three vocal parts.
Henry Cowell’s \textit{Mosaic Quartet} (1935) allowed performers to arrange its fragments in any order.
Morton Feldman’s \textit{Projection I} (1950)  introduced a novel form of graphical notation where tempo, timbre, and duration are indicated, but pitch and dynamics are left to the performer to decide.
John Cage also embraced chance in \textit{Music of Changes} (1951), using the \textit{I Ching} \cite{wilhelm2001ching} to guide compositional decisions.
In Earle Brown’s \textit{Twenty-Five Pages} (1953), up to 25 performers can determine page order and orientation.
Karlheinz Stockhausen’s \textit{Klavierstück XI} (1956) scattered 19 musical segments across a page, allowing performers to determine their order in real-time.
Iannis Xenakis introduced probabilistic models into composition. His \textit{Pithoprakta} (1956), meaning “actions through probability”, applies statistical mechanics of gases, and his book Formalized Music (1971) remains foundational in the field~\cite{xenakis1992formalized,harley2015iannis}.
Witold Lutosławski’s \textit{Venetian Games} (1961) employed aleatoric counterpoint, with timing and rhythmic coordination left to chance. Terry Riley’s \textit{In C} (1964) consists of short sequences that performers repeat as many times as desired before going on to the next.
Cornelius Cardew’s graphic scores in \textit{Treatise} (1967) allows for absolute interpretive freedom.
Louis Andriessen’s \textit{Worker’s Union} (1975) fixed rhythms but left pitches, number of instruments, and performance length to chance.
Brian Eno’s \textit{Music for Airports} (1978) redefined composition as a process rather than a product, using looping tape segments of varying lengths to create continuously evolving soundscapes \cite{Eno_Biography}. 
With the rise of electronic instruments and computers, new generative paradigms emerged. Modular synthesizers, \textit{e.g.}, enable creating self-generating music by patching signals that incorporate stochastic controls. Finally, current AI models are emerging as a promising generative approach, utilizing deep learning and probabilistic models to generate music.

In short, we introduced the history of generative music to note that modern generative AI 
has the potential to contribute to the already established field of generative music.

\subsection{Community}
\label{sec:festivals}

We also found that many AI music artists began their journey in `hackathons' (\textit{e.g.}, HAMR at ISMIR) or workshops held at festivals, such as CTM in Berlin, Gamma in St. Petersburg, Mutek in Montreal and Tokyo, or Sónar in Barcelona. AI artists also gathered at the Machine Learning for Creativity and Design Workshop at NeurIPS, around the AI Song Contest, and in Discord servers. 
These environments provide access to like-minded artists and technical mentorship, crucial for navigating the complex intersection of music and AI. Given that AI-based music creation can be technically demanding, musicians often collaborate with technical teams to manage these aspects. While user interfaces are improving, working with AI music  still requires a certain degree of technical proficiency.

\subsection{AI music tools}
\label{sec:musicAItools}

Beyond the artistic works we review, AI music tools are also shaping ongoing artistic trends. Though not the focus of our study, we include a brief summary for completeness.

Due to the computational demands of current AI models, many tools are hosted on cloud-based web platforms. Examples include full-song generation (like \textit{Stable Audio}, \textit{Suno}), source separation (\textit{AudioShake}, \textit{lalal.ai}), voice synthesis (\textit{ElevenLabs}), or lyrics generation (\textit{ChatGPT}).
AI plugins have also been introduced to work on DAWs and improve workflow efficiency.
Examples include MIDI generation (\textit{Magenta}, \textit{ORB}), mastering (\textit{LANDR}), AI-assisted mixing (\textit{iZotope}), or effects (\textit{Neutone}, \textit{Combobulator}). Finally, some AI music tools rely on dedicated hardware for tangible interactions. Examples include sample generators (\textit{NSynth Super}), Eurorack-compatible modules (\textit{Neurorack}), or AI keyboards (\textit{AWSDeepComposer}).

\section{Conclusions}
\label{sec:conclusions}

Most of the collected musical artworks use AI as a tool for co-composition or sound design, allowing artists to maintain creative control. Others, however, have explored the absence of artistic agency as part of their creative process. Furthermore, AI music often embraces AI's flaws and aesthetics, but at times it can be nearly indistinguishable from non-AI music. Some explored AI's imperfections and artifacts to evoke the uncanny. Others used AI's multi-genre generation capabilities to produce music in multiple (and potentially new) genres. Additionally, artists curate datasets to train their own models, similar to how producers program synthesizers to create their unique sounds. This example reveals the common technical challenges AI musicians face, a shared difficulty among those in the field. Innovative uses of AI include multilanguage song releases and exploring its potential in live performances and installations, bringing AI into formats like opera, soundtracks, ballets, and online installations. Another key innovation is using AI as an artistic medium, allowing generative AI albums to become a dynamic medium that evolves with each listen. AI voice releases are also an artistic medium that enables new interactions between artists and fans, raising interesting questions around authorship and digital identity.
We also discussed the importance of casual music creators and their role in the commodification of music creation (in a context where millions of tracks are being generated weekly \cite{UDIO}, with our study covering $\approx$ 300 artworks).
As AI music becomes more culturally relevant, public opinion is simultaneously shaping and influencing its development. 
Parallels with past narratives showcase that current trends and cultural pushback are not new.

We hope our work contributes to understanding how early adopters have used AI so far and serves as inspiration for future AI musicians, encouraging the creation of music that leaves a lasting, positive impact on music history.

\subsection{Limitations}
\label{sec:limitations}

As art is understood differently across time and cultures, our study is limited by our time (2025) and place (West). Further, our review captures trends based on existing data but does not reflect future shifts in AI music. Our study captures a moment in AI music, not its full trajectory.

We collected a large set of AI music works for our qualitative and quantitative analysis. 
But collecting this corpus was challenging, as there is no centralized database of AI music artists. Hence, the process is not strictly systematic. 

Although we discussed the importance of the casual creator and the commodification of music, analyzing the vast amount of AI music by casual creators is beyond our scope.

The authors speak English, Catalan, French, and Spanish, which limited our access to music scenes in Asia and Africa. We encourage further research to explore these underrepresented regions and challenge our findings. 

Our discussion section is limited to a few historical examples for contextualizing AI music artworks, as running a review of historical examples is also outside our scope.

\bibliography{ISMIRtemplate}

\end{document}